\documentclass[runningheads]{cl2emult}

\usepackage{makeidx}  % allows index generation
\usepackage{graphicx} % standard LaTeX graphics tool
\usepackage{subeqnar} % subnumbers individual equations
\usepackage{multicol} % used for the two-column index
\usepackage{cropmark} % cropmarks for pages without
\usepackage{eso}      % placeholder for figures
\makeindex            % used for the subject index
\begin{document}

\title*{The XMM/Megacam-VST/VIRMOS\\ Large Scale Structure
Survey}
%

%\vspace{-1cm}  Talk given at ****
 \toctitle{The
XMM/Megacam-VST/VIRMOS \protect\newline Large Scale Structure
Survey}
%\footnote{AAAAAAAAAAA}
% allows explicit linebreak for the table of content
%
%
\titlerunning{The XMM-LSS Survey}
% allows abbreviation of title, if the full title is too long
% to fit in the running head
%
\author{Marguerite Pierre\inst{1}
\and The XMM-LSS Consortium \inst{2}}

\authorrunning{M. Pierre}
% if there are more than two authors,
% please abbreviate author list for running head
%
%
\institute{CEA Saclay, Service d'Astrophysique\\
     F-91191 Gif sur Yvette, France\\
     mpierre@cea.fr
\and
http://vela.astro.ulg.ac.be/themes/spatial/xmm/LSS/cons\_e.html}

\maketitle              % typesets the title of the contribution

\vspace{-6.5cm}
\noindent
%\hspace{2cm}
\centerline{\tt Talk given
at {\sl Mining The Sky}}\\
 \centerline{\tt A joint MPA/ESO/MPE conference, Garching, 7/31 - 8/4, 2000}
 \vspace{5.5cm}

\begin{abstract}
The objective of the XMM-LSS Survey is to map the large scale
structure of the universe, as highlighted by clusters and groups
of galaxies, out to a redshift of about 1, over a single $8\times
8$ sq.deg. area. For the first time,  this will reveal the
topology of the distribution of the deep potential wells  and
provide statistical measurements at truly cosmological distances.
In addition, clusters identified via their X-ray properties will
form the basis for the first uniformly-selected, multi-wavelength
survey of the evolution of clusters and individual cluster
galaxies as a function of redshift. The survey will also address
the very important question of the QSO distribution within the
cosmic web.
\end{abstract}

\section{Context}
As the largest gravitationally bound entities, clusters of
galaxies play a key role in our understanding of the universe. In
particular, the redshift evolution of both their individual
properties and global space distribution are essential to
constrain cosmological scenarios. Since clusters originate from
high amplitude initial density fluctuations, they are rare events
and dedicated search programmes are necessary to provide
homogeneous samples suitable for statistical studies (the mean
cluster/group number density\footnote{We assume H$_{o}$ = 50
km/s/Mpc  and q$_{o}$ = 0.5 throughout this paper} is of the order
5~$10^{-6}$ Mpc$^{-3}$ \cite{ebe}). A notable sample is the Abell
(ACO) catalogue \cite{abe} which enabled the first measurement of
the local cluster power spectrum \cite{ein}. Optical catalogues
are, however, severely hampered by projection effects and galaxy
density contrasts with respect to the background become marginal
beyond $z\sim 1$, unless detailed multi-color information is
available, together with sophisticated detection algorithms. In
this context, the X-ray wave-band represents much more than a
useful alternative: it is a secure and straightforward approach. A
high latitude galactic field observed at medium sensitivity ($\sim
10^{14}$ erg/s/cm$^{2}$) shows basically two types of objects: QSO
(pointlike) and clusters (extended), the cluster X-ray emission
being due to the hot diffuse gas trapped in the cluster potential.
Moreover, the X-ray temperature and luminosity can be related to
the cluster total mass, provided the physics of the intra-cluster
medium (ICM) is properly modelled. From ROSAT and ASCA
observations, the current status of the X-ray cluster research can
be summarized as follows: there is no significant evolution in the
cluster luminosity function out to $z\sim 0.8$ \cite{ros}, or in
the $L_{X}-T_{X}$ relationship out to $z\sim 0.5$ \cite{mus}; the
power spectrum of the {\em local} X-ray cluster population is
remarkably similar to that of galaxies, with a higher scaling
\cite{guz}.\\ Due to its unrivalled sensitivity (Fig. 1), its
large field of view (30') and good PSF (FWHM = 6" on-axis), XMM
opens a new era for cluster studies, and will not be superseded
for many years to come.

\begin{figure}
\centering
\includegraphics[width=6cm,angle=-90]{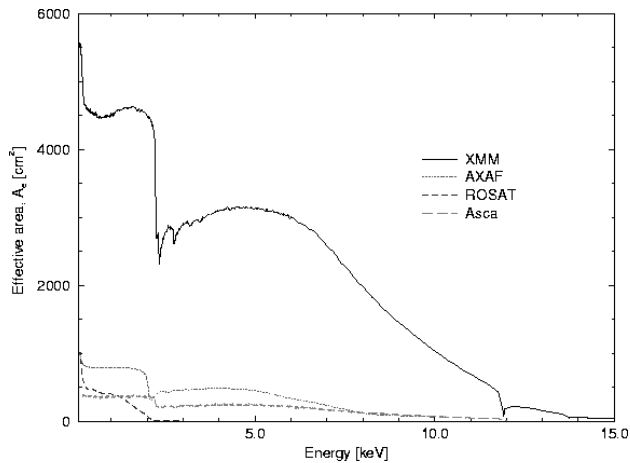}
\caption[]{The XMM effective area as a function of energy compared
to other X-ray satellites. (AXAF = {\sl Chandra})} \label{eps1}
\end{figure}

 We have, thus, designed an
XMM wide area survey with the aim of tracing the large scale
structure of the universe out to a redshift of $z\sim 1$, as
underlined by clusters and QSOs: {\sc The XMM Large Scale
Structure Survey (XMM-LSS)} (Fig. 2).

\begin{figure}
\centering
\includegraphics[width=6cm,angle=-90]{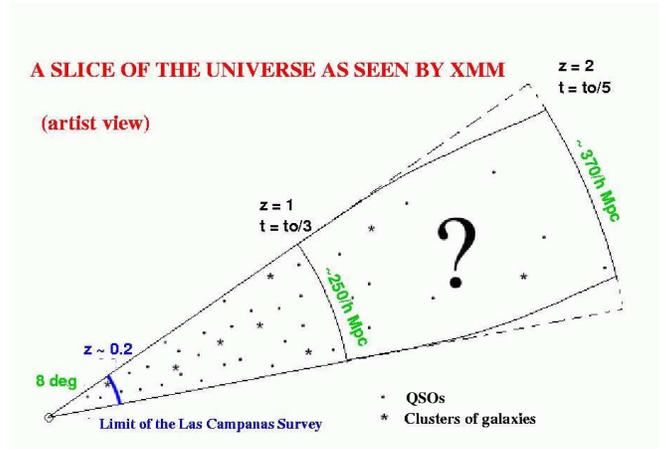}
\caption[]{An artist view of the XMM-LSS. Transversal distances
are in comoving units. QSOs should be discovered out to a redshift
of $\sim 4$. Some  300 sources  per sq.deg. are expected, out of
which about one tenth will be galaxy clusters. {\bf For the first
time, a huge volume of the distant universe will be uniformly
sampled}.} \label{eps2}
\end{figure}

The wide scope of the project has motivated the set-up of a large
consortium in order to facilitate both the data
reduction/management and the scientific analysis of the survey.
The XMM-LSS Consortium comprises the following institutes: Saclay
(Principal Investigator), Birmingham, Bristol, Copenhagen, Dublin,
ESO/Santiago, Leiden, Liege, Marseille (LAM), Milan (AOB), Milan
(IFCTR), Munich (MPA), Munich (MPE), Paris(IAP), Santiago (PUC).

\section{X-ray and follow-up observations }
 \underline{The
survey design}\\ The survey consists of adjacent 10 ks XMM
pointings and will cover a region of $8\times 8$ sq.deg. (with a
deeper central 2 sq.deg. area); the mean sensitivity will be about
$5~10^{-15}$ erg/s/cm$^{2}$ in the [0.5-2] keV band. It is located
around RA = 2h20, Dec = -5deg. A source density of $\sim 300$
 per sq.deg. is expected including: 65\% QSO; 15\% nearby galaxies;
12\% galaxy clusters; 8 \% stars. The histogram of the predicted
cluster redshift distribution is shown on Fig. 3.

\begin{figure}
\centering
\includegraphics[width=3cm,angle=-90]{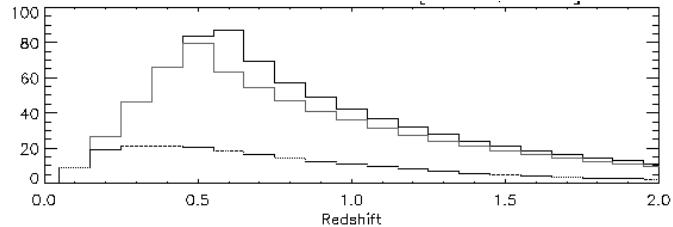}
\caption[]{The predicted XMM-LSS cluster redshift distribution,
computed using the local cluster luminosity function and
properties; redshifted thermal spectra convolved with the XMM
response were simulated, source number counts computed and finally
compared to the survey sensitivity limit. Three detection bands
are shown ([2-10], [0.6-8]and [0.4-4] keV, from bottom to top
respectively). The [0.4-4]
 keV band is the most sensitive for
clusters, whereas the hardest one is quite inefficient since the
majority of the cluster/group population has a temperature of the
order of 2-3 keV (restframe). Up to 800 clusters are expected out
to $z = 1$ and of the order of 100 between $1<z<2$ (if there is no
evolution). } \label{eps2}
\end{figure}

\noindent
 \underline{Basic follow-up}\\ In order to ensure the
necessary identification and redshift measurement of the X-ray
sources, we have started an extensive multi-wavelength follow-up
programme.
 Optical and NIR imaging
has been initiated at CFHT and CTIO and will be then uniformly
performed by the 2nd generation of wide field imagers such as
Megacam/WFIR (CFHT) and VST (ESO). Subsequent spectroscopic
identifications and redshift measurements will be performed by the
VLT/VIRMOS instrument and other 4-8m class telescopes to which the
consortium has access. Entire coverage of the region by the VLA is
underway at 90 and 400 cm.

%\pagebreak

\section{Expected science} The XMM-LSS has been designed such as to {\bf
enable, for the first time, the determination of the cluster
2-point correlation function in two redshift bins
($0<z<0.5,~0.5<z<1$)}, with an accuracy better than 10\% for the
correlation length. Considered on a more qualitative (topological)
point of view, we shall obtain a 3D map of the deep potential
wells of the universe within an unprecedented volume. Both aspects
will have profound cosmological implications. Beside this main
goal, thanks to the unique data set to be collected,  several
other fundamental aspects will be addressed.\\
 - First of all, as apparent on Fig.
3, we shall be in a position to test the existence of massive
clusters out to a redshift of $\sim 2$. This is also of key
importance for constraining cosmological scenarios.\\ - We shall
compute, to a high degree of accuracy, the QSO 2-point correlation
function out $z \sim 4$. \\
 - The study of the combined
X-ray/optical/radio evolution of clusters and QSOs, of their
galaxy content and of their environment is an obvious
``by-product" of the XMM-LSS. This aspect is to be especially
important at redshifts beyond 1, where merger and star formation
are expected to be significantly more active than in the local
universe. Indeed, preheating and shocks are thought to influence
the ICM properties of forming clusters, i.e. before they reach a
relaxed state. Moreover,  these effects  are redshift dependent
since cluster sizes, densities and temperatures are expected to
vary as a function of redshift, on a purely gravitational basis.
Although there is both theoretical and observational evidence for
traces of feedback in the low redshift cluster population
\cite{dav} \& \cite{met}, its influence needs to be assessed and
quantified at earlier times  \cite{men}. The radio data will
provide an important source of complementary information for our
understanding of merger processes, as well as the presence of
energetic particles and magnetic fields which are likely to also
affect the state of the ICM. \\ - Finally, it will be possible to
see how the QSO population fit into the LSS network defined by the
cluster/group population. This ``external view'' of the QSOs is a
fundamental complement to the ``internal view'', i.e. the unified
AGN scheme; indeed, this latter approach alone neither explains
the observed strong QSO clustering, nor the fact that BL Lac
objects, for instance, are preferentially found in clusters or
groups \cite{wur}. The environmental properties of AGNs is thus
crucial for the understanding of their formation (mergers, initial
density perturbations of a peculiar type, etc.). The XMM-LSS data
set will also provide decisive statistical information regarding
the effect of gravitational lensing on QSO properties.\\
\underline{Advanced follow-up}\\ Subsequently to the core
programme science, further detailed follow-up will be undertaken
for objects that appear as especially relevant. For instance, deep
XMM pointings will be used to study high-$z$ forming cluster
complexes \cite{pie}. Also, the expected high density of QSOs in
the survey may form the basis of high-resolution optical
spectroscopy within a sub-area, in order to map the
L$\alpha$-forest and, thus, obtain a detailed 3D picture of the
structures where most of the baryons are expected to be located
\cite{cen}.\\ The deep and high quality optical coverage of the
entire 64 sq.deg. area by Megacam will enable an unprecedented
weak-lensing analysis \cite{meg}. Its cosmological implications
will be directly compared to the constraints derived from the
XMM-LSS cluster sample.\\ Finally, Sunyaev-Zel'dovich observations
(S-Z) are also foreseen. In a first step, individual XMM-LSS
clusters will be observed; together with the X-ray, optical and
radio observations, this will enable a truly statistical analysis
of the physics of the ICM as a function of redshift. On the long
term, S-Z  mapping of part or of the entire XMM-LSS area should
provide invaluable information on the low density structures such
as cluster outskirts as well as their connections to the cosmic
filaments.

\section{Simulations}
We illustrate, by two examples,  the characteristics of the
XMM-LSS (Fig. 4 \& Fig. 5). The captions outline some of the major
impacts of the project.\\

\noindent

\begin{figure}
\underline{The 3D cluster distribution}\\ \centerline{
\includegraphics[width=6cm,angle=0]{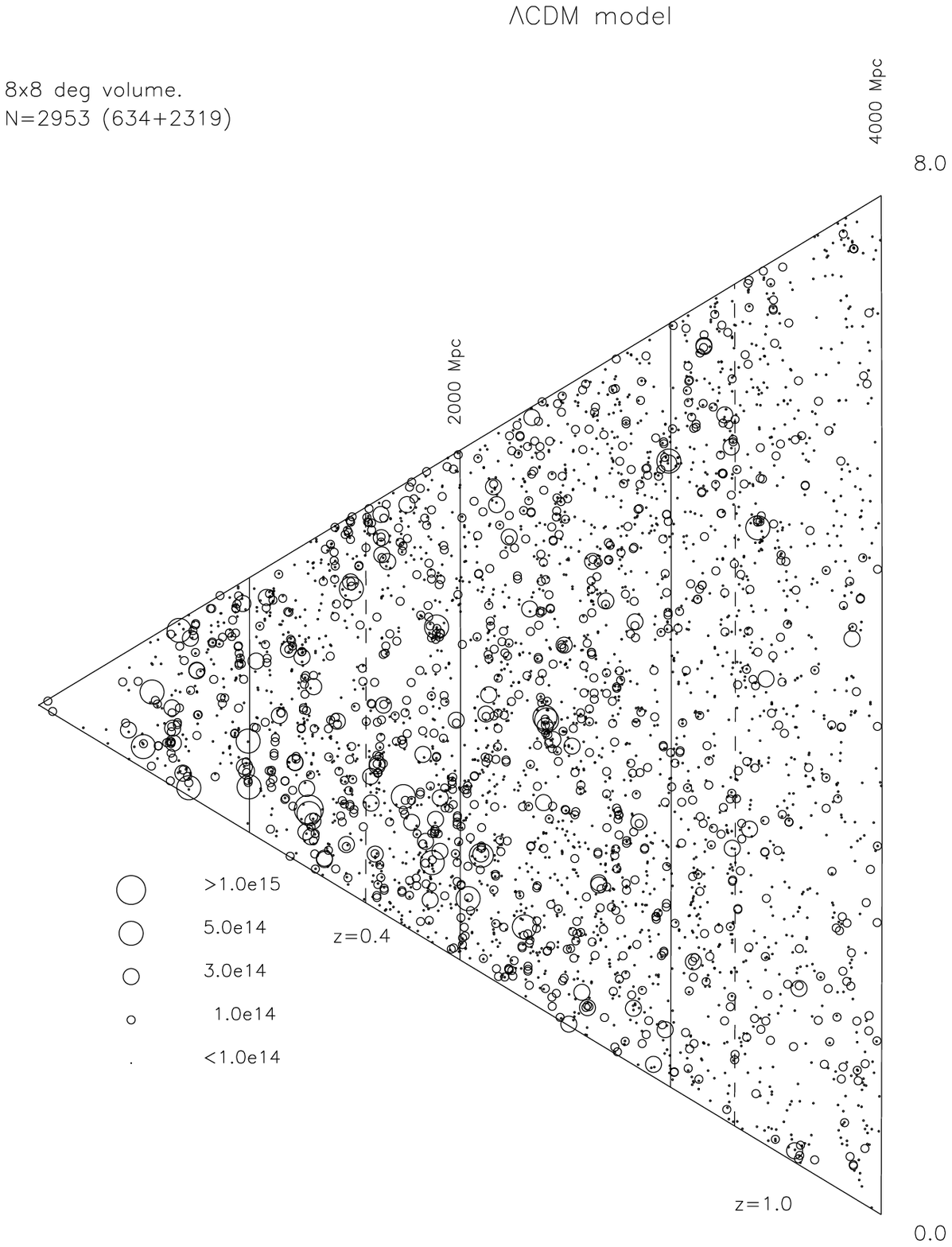}
\includegraphics[width=6cm,angle=0]{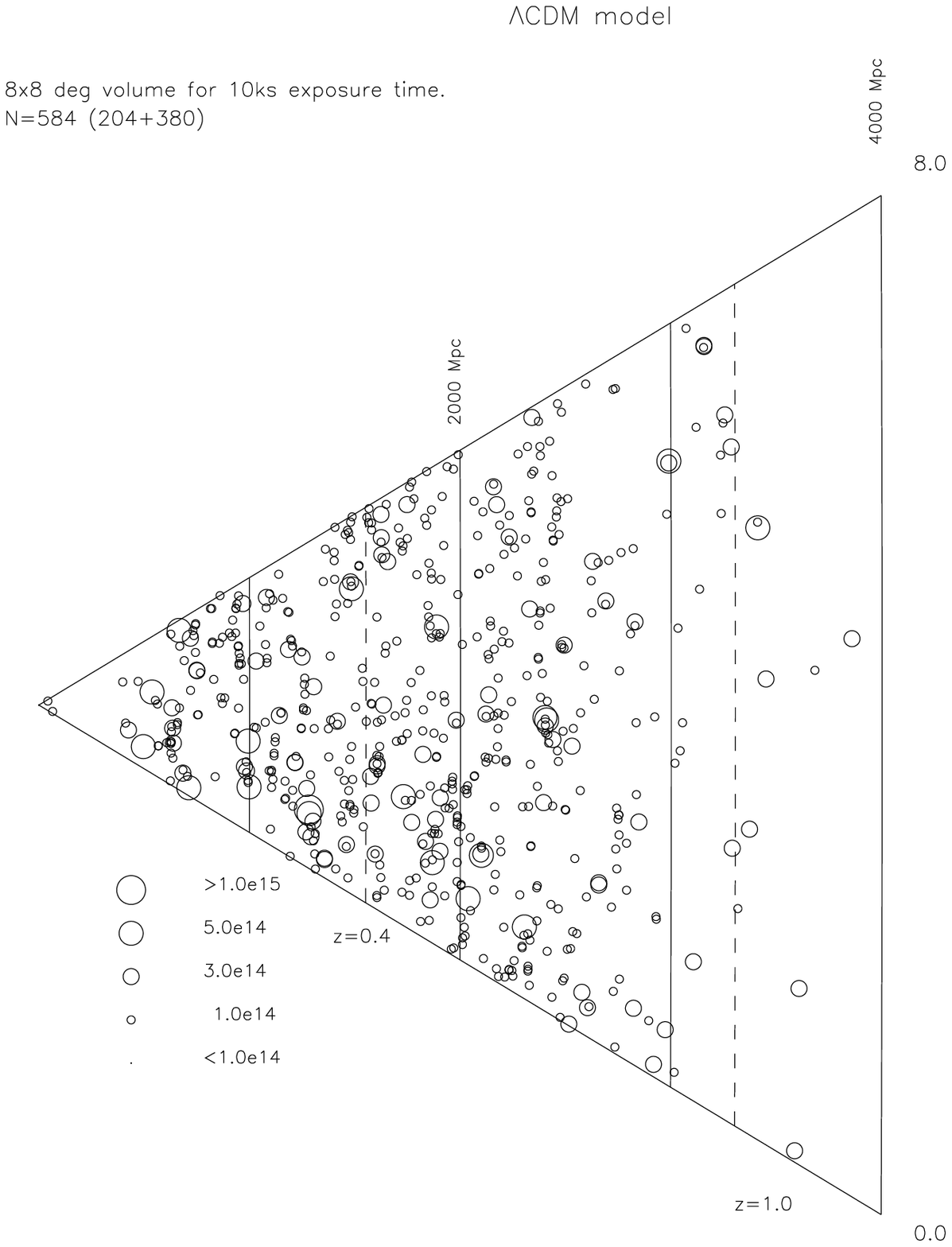}}
\caption[]{Simulation of the XMM-LSS cone, using the Hubble Volume
Lightcone cluster catalogue for a $\Lambda$CDM model \cite{hub}.
Symbol sizes indicate cluster masses. Together with Fig. 2, this
wedge diagram shows, in a striking manner, how {\bf the XMM-LSS
will provide the next hierarchical step as compared to traditional
galaxy surveys}. Points are now galaxy clusters  which are the
carriers of a cosmologically significant parameter: their mass.
Predicted numbers of clusters in the $0<z<0.5, 0.5<z<1$ bins are
given in brackets. {\em Left:} the cluster distribution; cosmic
evolution can be appreciated from the decrease of the number
density  of massive clusters at high redshift. {\em Right:}
convolution by the XMM-LSS selection function: only massive
clusters are detectable a high redshift. }\label{eps2}
\end{figure}

\begin{figure}
\underline{An XMM-LSS field}\\
%\centerline{\includegraphics[width=11cm,angle=0]{pierre5.ps}}
\caption[]{Simulation of a 10 ks XMM-LSS field, encompassing a
cosmic filament at  $z = 0.5$, whose properties have been
estimated from high resolution hydrodynamical simulations. {\
Left:} The filament  photon image alone. Three galaxy groups are
conspicuous (masses of 1.7, 3.2, 3.5 $10^{14}$ M$_\odot$), but not
the diffuse filamentary medium linking the collapsed objects. {\em
Right:} same field with the back/foreground QSO population now
added; the image has been filtered using a multi-resolution
wavelet algorithm. The groups clearly show up as diffuse objects.
In the XMM-LSS Survey, it will be possible to infer the  existence
of cosmic filaments through the presence of chains of groups and
clusters; then, subsequent weak-lensing analysis will probe the
gravitational properties of the underlying dark matter. Details on
the XMM simulations of the cosmic network can be found in
\cite{pie}. } \label{eps2}
\end{figure}

\section{Conclusion}

The ultimate goal of the XMM-LSS survey is to map the matter
distribution out to $z = 1$ over a $8 \times 8$ sq.deg. area,
using three different methods\footnote{A detailed description of
the XMM-LSS project  (consortium, multi-$\lambda$ follow-up, data
management and analysis, status of the observations) can be found
at
http://vela.astro.ulg.ac.be/themes/spatial/xmm/LSS/index\_e.html}:
\\ - X-ray clusters
and QSOs\\
- weak-lensing analysis \\ - Sunyaez-Zel'dovich effect.

\end{document}